\documentclass[doublecol]{epl2} 

\usepackage{amssymb,amsmath,color,ulem}

\title{How low can SUSY go? \\ Matching, monojets and compressed spectra.}

\shorttitle{How low can SUSY go?} 

\author{Herbi K. Dreiner\inst{1} \and Michael Kr\"amer\inst{2} \and Jamie Tattersall\inst{1}}
\shortauthor{Herbi K. Dreiner \etal}

\institute{                    
  \inst{1} Universit\"{a}t Bonn, Physikalisches Institut, Nu\ss allee 12, 53115 Bonn, Germany \\
  \inst{2} Institute for Theoretical Particle Physics and Cosmology, RWTH Aachen, D-52056 Aachen, Germany
}
\pacs{14.80.Ly}{Supersymmetric partners of known particles}
\pacs{12.60.Jv}{Supersymmetric models}

\abstract{
If supersymmetry (SUSY) has a compressed spectrum then the current mass limits from the LHC can be 
drastically reduced. We consider a possible \textquoteleft worst case' scenario where the gluino and/or squarks are 
degenerate with the lightest SUSY particle (LSP). The most sensitive searches for these compressed 
spectra are via the final state LSPs recoiling against initial state radiation (ISR). Therefore it is vital that the ISR is 
understood and possible uncertainties in the predictions are evaluated. We use both MLM (with Pythia 6) and CKKW-
L (with Pythia 8) matching and vary matching scales and parton shower properties to accurately  determine the 
theoretical uncertainties in the kinematic distributions. All current LHC SUSY and monojet analyses are employed and we find the most constraining limits come from the CMS  Razor and CMS monojet searches. For a scenario of 
squarks degenerate with the LSP and decoupled gluinos we find $M_{\tilde{q}}>340$~GeV. For gluinos degenerate with the 
LSP and decoupled squarks, $M_{\tilde{g}}>500$~GeV.  For equal mass squarks and gluinos degenerate with the 
LSP, $M_{\tilde{q},\tilde{g}}>650$~GeV. \\ \\ BONN-TH-2012-18}

\begin{document}

\maketitle

\section{Introduction}
\label{sec:intro}

Supersymmetry (SUSY) \cite{Martin:1997ns} has now been extensively searched for at the LHC \cite{ATLAS-
CONF-2011-096,CMS-Mono-B,ATLAS-CONF-2012-033,Chatrchyan:2011zy,CMS-PAS-SUS-11-004-B,CMS-PAS-
SUS-12-002-B,CMS-PAS-SUS-12-005-B}. In the limit that the lightest supersymmetric particle is massless the bounds on 
equal mass squarks and gluinos are $\gtrsim1.5$~TeV. However, if the LSP is massive and degenerate with the squarks and/or gluinos, these limits can be severely weakened \cite{Alwall:2008ve,Alwall:2008va,LeCompte:2011cn,LeCompte:2011fh,Izaguirre:2010nj}. 

In the case of \textquoteleft extreme' compression, initial state radiation (ISR) in the form of jets may be used to discover the 
model \cite{Gunion:1999jr}. This method has now been used to set mass limits on squarks and gluinos 
\cite{LeCompte:2011cn,LeCompte:2011fh} and stops \cite{He:2011tp,Drees:2012dd}. Other ideas to look for compressed 
SUSY include using monophotons \cite{Belanger:2012mk} and soft leptons \cite{Rolbiecki:2012gn}. Here we look at 
simplified models in which the first two generations of squarks and/or the gluinos can be quasi-degenerate with  the 
LSP and set limits on the particle masses. These models represent a \textquoteleft worst case' scenario for R-parity 
conserving SUSY at the LHC and thus give a model independent bound on these masses. 

The limits completely rely on the accuracy of the ISR prediction for radiated jets. Thus, we use two independent matching 
schemes (CKKW-L \cite{Catani:2001cc,Lonnblad:2001iq}) and (MLM \cite{Mangano:2006rw}) to carefully understand 
the accuracy of the prediction. In addition, matching scales and parton shower variables are changed to ensure that the limits 
are robust.

\section{ISR and matching procedure}
\label{sec:ISR}

\begin{figure*}[ht!]
  \begin{center}
    \includegraphics[scale=0.45]{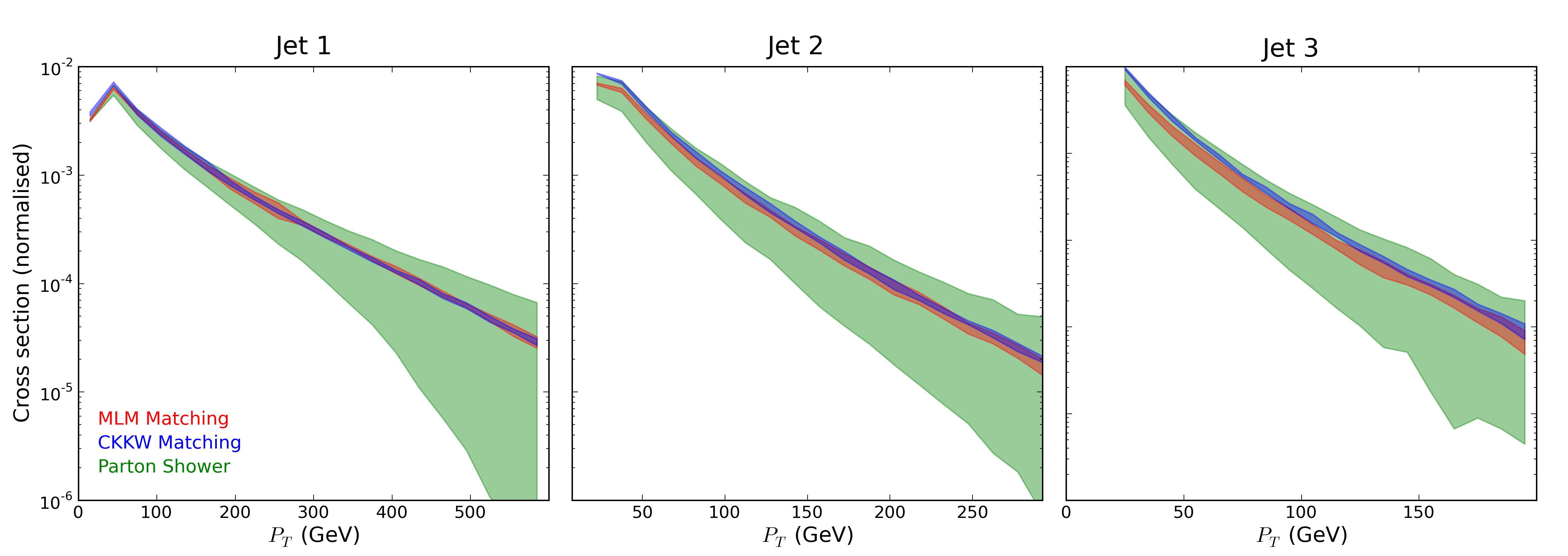} \hspace{0.8cm}
  \end{center}\vspace*{-0.4cm}
  \caption{A comparison of the uncertainty in the ISR jet $p_T$ distribution for the production of squarks ($M_{\tilde{q}}
  =500$~GeV) between the parton shower prediction (green, light), MLM matching (pink, medium) and CKKW matching (blue, dark). The parton 
  shower uncertainty is found by varying the Pythia 6 and 8 parton showers between the \textquoteleft wimpy' and 
  \textquoteleft  power' settings \cite{Plehn:2005cq}. The matching uncertainties are found in both cases by varying the 
  matching scales between 50 and 200~GeV and additionally for MLM matching by varying the parton shower between the 
  \textquoteleft wimpy' and \textquoteleft power' settings. For the softer jet 3 (the first unmatched jet), the relative uncertainty of the parton shower approach is reduced since the phase space for this emission is better constrained.  \label{fig:MGvsPyth}}
\end{figure*}

To reliably set limits in SUSY models with compressed spectra it is vital that the ISR is well modeled and any uncertainties 
are understood. To achieve this we must use a matrix element prediction for hard and well separated partons 
whilst using a parton shower for soft and/or collinear jets. To bring the two approaches together we must match the two 
methods in a consistent algorithm. 

The reason that a matrix element prediction for hard radiation is required is that a parton shower approach contains large 
uncertainties in this regime. They are mainly due to the choice of starting scale for parton 
shower evolution. This has been shown to have a large effect in squark and gluino production 
\cite{Plehn:2005cq} and we show the variation between the \textquoteleft wimpy' and \textquoteleft power' shower settings in Pythia 6 \cite{Sjostrand:2006za} and 8  \cite{Sjostrand:2007gs} in Fig.\ref{fig:MGvsPyth}.

In order to match the matrix element to the parton shower we must be careful to avoid double counting so that areas of phase 
space are only filled by one approach. In addition we would like a smooth transfer between the different areas of validity. 
Finally, the prediction should not have a significant dependence on the chosen matching scale or 
parton shower. Within SUSY we have the additional problem that we can double count events with 
resonant propagators, which must be removed  \cite{Plehn:2005cq,Alwall:2008qv}.

To independently check the predictions of the matching algorithm, two approaches were used. First is the integrated MLM 
\cite{Mangano:2006rw} matching available in Madgraph 5 \cite{Alwall:2008qv,Alwall:2011uj} which is interfaced with the Pythia 
6 shower \cite{Sjostrand:2006za}. To estimate the uncertainty, we varied the matching scale between 50 
and 200~GeV and independently the parton shower between the \textquoteleft wimpy' and \textquoteleft 
power' settings. As seen in Fig.\ref{fig:MGvsPyth}, this results in a significant reduction in the uncertainty 
compared to the parton shower alone.
The second approach was the CKKW-L \cite{Catani:2001cc,Lonnblad:2001iq} matching algorithm, developed for Pythia 8 
\cite{Sjostrand:2007gs,Lonnblad:2011xx}. It was adapted to work with SUSY\footnote{We would especially like 
to thank Stefan Prestel for his invaluable help in adapting the algorithm.} and gives consistent results to those obtained with 
MLM matching, Fig.\ref{fig:MGvsPyth}.

\section{Simplified Models}
\label{sec:SimpMod}

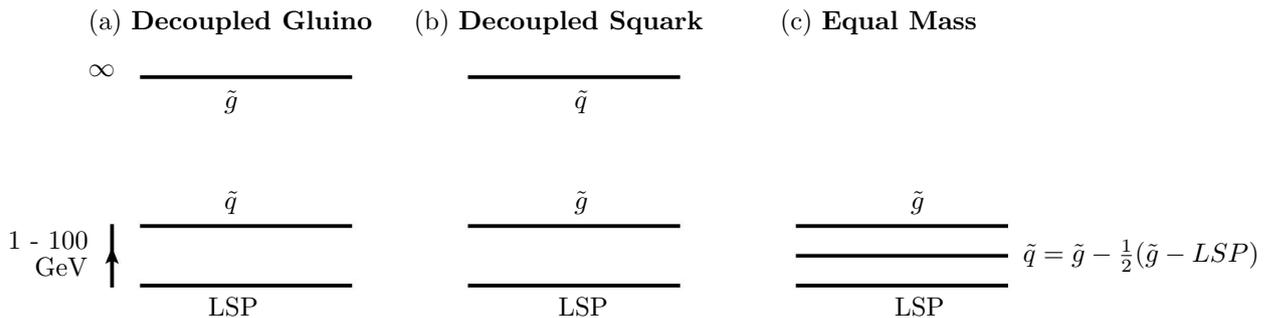
\begin{figure*}[ht!]
  \begin{center}
    \includegraphics[scale=0.25]{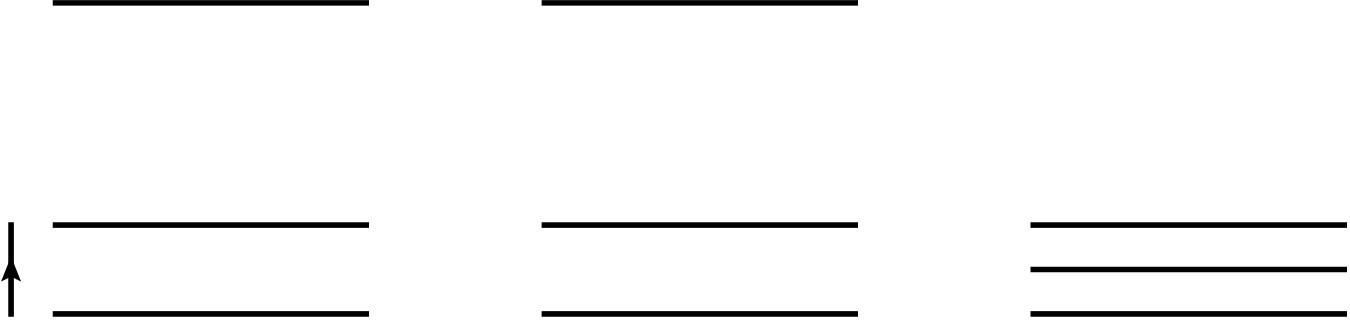}
	   \put(-345,98){(a) \bf{Decoupled Gluino}}
	  \put(-223,98){(b) \bf{Decoupled Squark}}
	    \put(-86,98){(c) \bf{Equal Mass}}
	      \put(-294,68){$\tilde{g}$}
	      \put(-163,68){$\tilde{q}$}
	      \put(-345,80){$\infty$}
	      \put(-294,30){$\tilde{q}$}
	      \put(-163,30){$\tilde{g}$}
	      \put(-37,30){$\tilde{g}$} 
		\put(5,10){$\tilde{q}=\tilde{g}-\frac{1}{2}(\tilde{g}-LSP)$}
	      \put(-300,-10){LSP}
	      \put(-169,-10){LSP}
	      \put(-43,-10){LSP}
	      \put(-375,15){1 - 100}
	      \put(-365,5){GeV}
  \end{center} 
  \caption{The spectra for the simplified models studied in this paper. For the \textquoteleft Decoupled Gluino' scenario we remove the gluino from the spectrum and vary the mass difference between the squarks and the LSP from 1 to 100~GeV. In the \textquoteleft Decoupled Squark' scenario we remove any squarks from the spectrum and vary the mass difference between the gluino and the LSP from 1 to 100~GeV (see accompanying text for caveats). In the equal mass scenario, the gluino is set as the most massive particle and the squark mass is halfway between the gluino mass and LSP mass. In all scenarios the squarks are a summation over the first two generations and the third generation are ignored. \label{fig:Spectrum}}
\end{figure*}

In order to reduce the SUSY parameter space and place model independent limits, we use three simplified models. 
The idea is to investigate the \textquoteleft worst case' scenario for R-parity conserving SUSY. We thus assume
 that either the first two generations of squarks or the gluinos or both are quasi-degenerate with the 
LSP. The degeneracy has the effect of making all of the SUSY decays invisible to the detector as the produced charged particles are too soft to be reconstructed. Therefore, events with ISR are solely relied upon to set any 
limits on the model.

Our first scenario is labeled the \textquoteleft Decoupled Gluino' model, Fig.\ref{fig:Spectrum}(a). Here the 
first two generations of squarks are quasi-degenerate with the LSP (1-100~GeV mass splitting) while the gluino is completely 
removed from the scenario. The idea is to set a lower limit on the first two generation squarks masses that is completely 
independent of the gluino mass.
The third generation of squarks are left free (obviously heavier than the LSP) because in general the Yukawa 
contribution to the running of the mass leads to a splitting between these and the first two
generations of squarks. However, a degenerate contribution can easily be added by simply rescaling the cross-section by 5/4 for 
only sbottoms or 6/4 for stops as well.

The second scenario we name the \textquoteleft Decoupled Squark' model, Fig.\ref{fig:Spectrum}(b). Here the gluino is quasi-degenerate with the LSP (1-100~GeV mass splitting) and the first two generations of squarks are removed from the model.
In the limit that all squarks are removed from the scenario it must be stated that the gluino becomes stable and a distinctive signal 
would therefore be seen as so called \textquoteleft R-hadrons'. In fact, even for moderate squark masses it is easy to make a 
gluino in a compressed spectra long-lived. However, it is possible that the third generation squarks could be much lighter than the 
other squarks. These could mediate prompt gluino decay whilst having a negligible impact on the search. Therefore we assume a 
prompt decaying gluino in this scenario as an interesting limiting case.\footnote{Such models are already being 
investigated by the LHC collaborations. \cite{CMS-PAS-SUS-11-016-B}}

As a third scenario we consider the \textquoteleft Equal Mass' model, Fig.~\ref{fig:Spectrum}(c). Here, the gluino mass 
is set quasi-degenerate with the LSP (1-100~GeV mass splitting) and the first two generations of squarks have a 
mass halfway between the gluino and the LSP,
\begin{equation}
 M_{\tilde{q}}=\frac{1}{2}(M_{\tilde{g}}+M_{LSP}).
\end{equation}
 The third generation squarks are once again ignored but there is no conceptual reason why they could not be added. However, they would only give a very small contribution to the final cross-section so our limit would remain practically unchanged.

Additional assumptions have to be made on the models in order to use the LHC analyses. Firstly, we assume that all decays are prompt and we have no new heavy states traversing the detector or producing displaced vertices. In any case, these signals are currently searched for and should provide a distinctive signature \cite{Chatrchyan:2012sp}. 
Secondly, when the mass splittings between the squarks/gluinos and the LSP is increased, we assume that the decay occurs directly to the next lightest particle in our decay chain. Therefore, we assume that no other states exist in between. In the limit of degeneracy, even if new states did exist, the phenomenology is unchanged because all the momentum of the initial parent particle is still transfered to the LSP. For increased splitting it is possible that the limits may be changed with multiple decays to many soft particles. However, we do not consider this possibility.

\section{Searches}
\label{sec:searches}

\begin{figure*}[ht!]
\begin{center}
\includegraphics[scale=0.4]{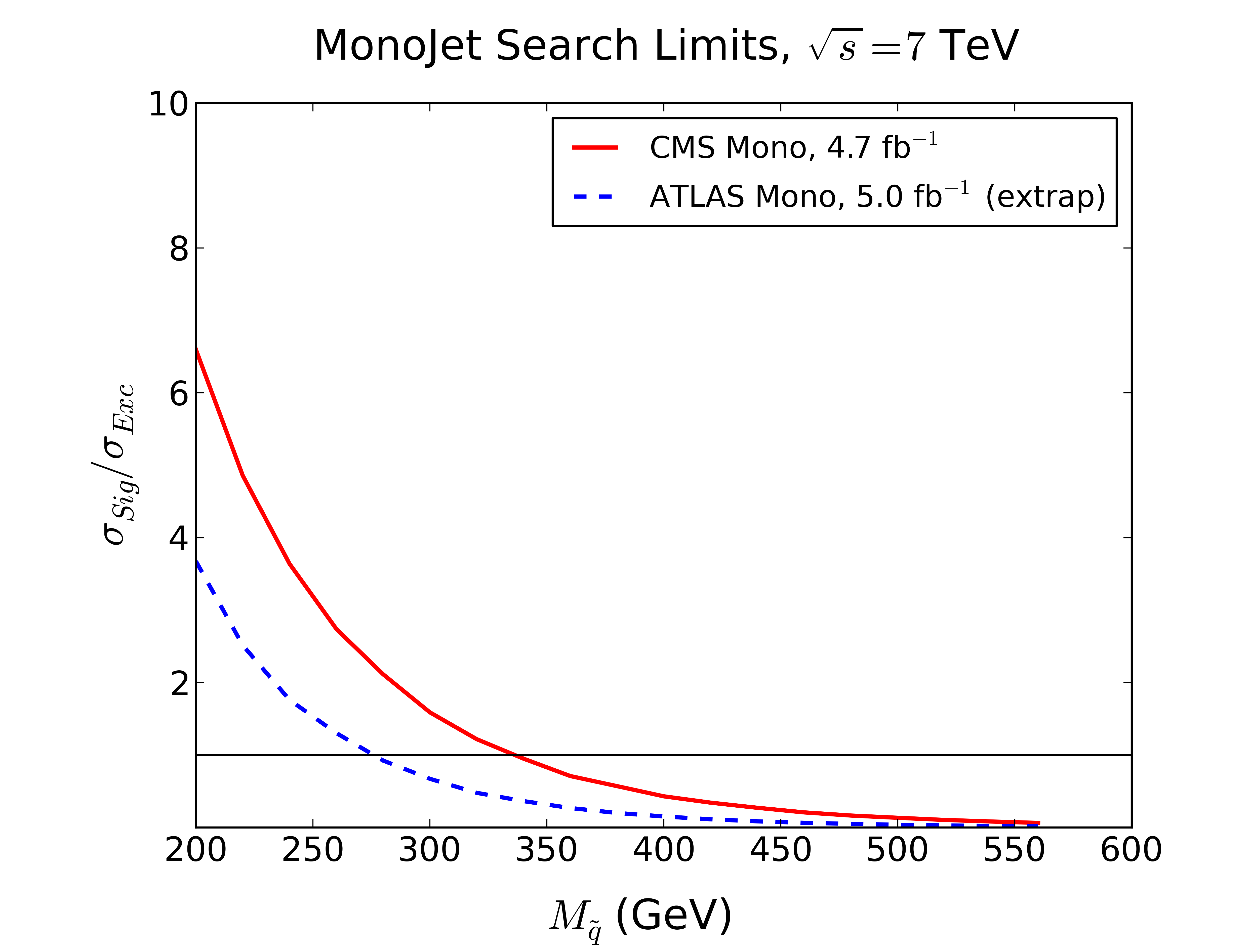}  
\includegraphics[scale=0.4]{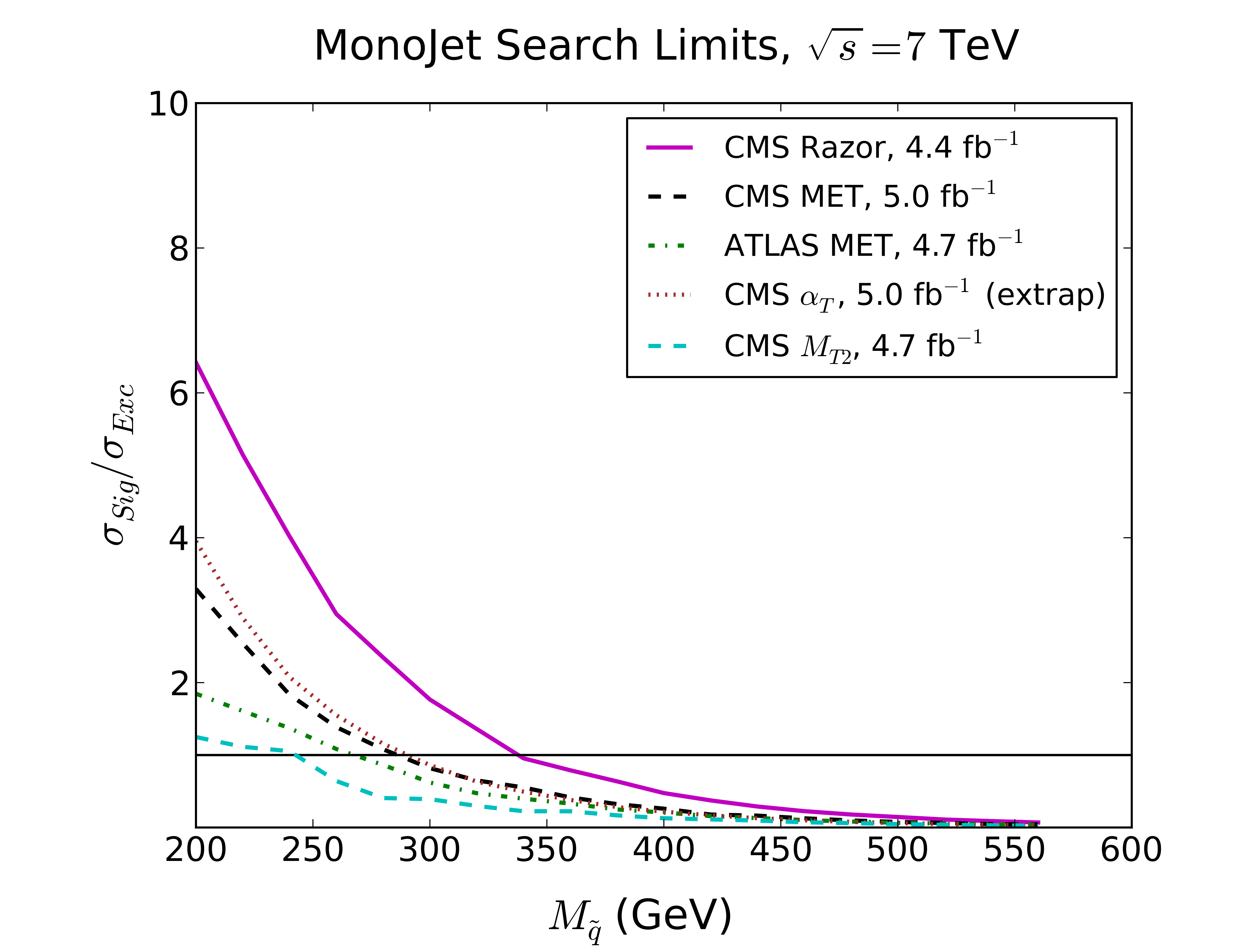}
\end{center} \vspace*{-0.4cm}
\caption{Ratios of the signal cross-section $(\sigma_{Sig})$ to the cross-section required for exclusion $(\sigma_{Exc})$ at the 95\% confidence level for squark masses in the decoupled gluino scenario, Fig.\ref{fig:Spectrum}(a). The monojet searches (left) and SUSY searches (right) are shown.\label{fig:SquarkLineLimit}} 
\end{figure*}

\begin{figure*}[ht!]
\begin{center}
\includegraphics[scale=0.4]{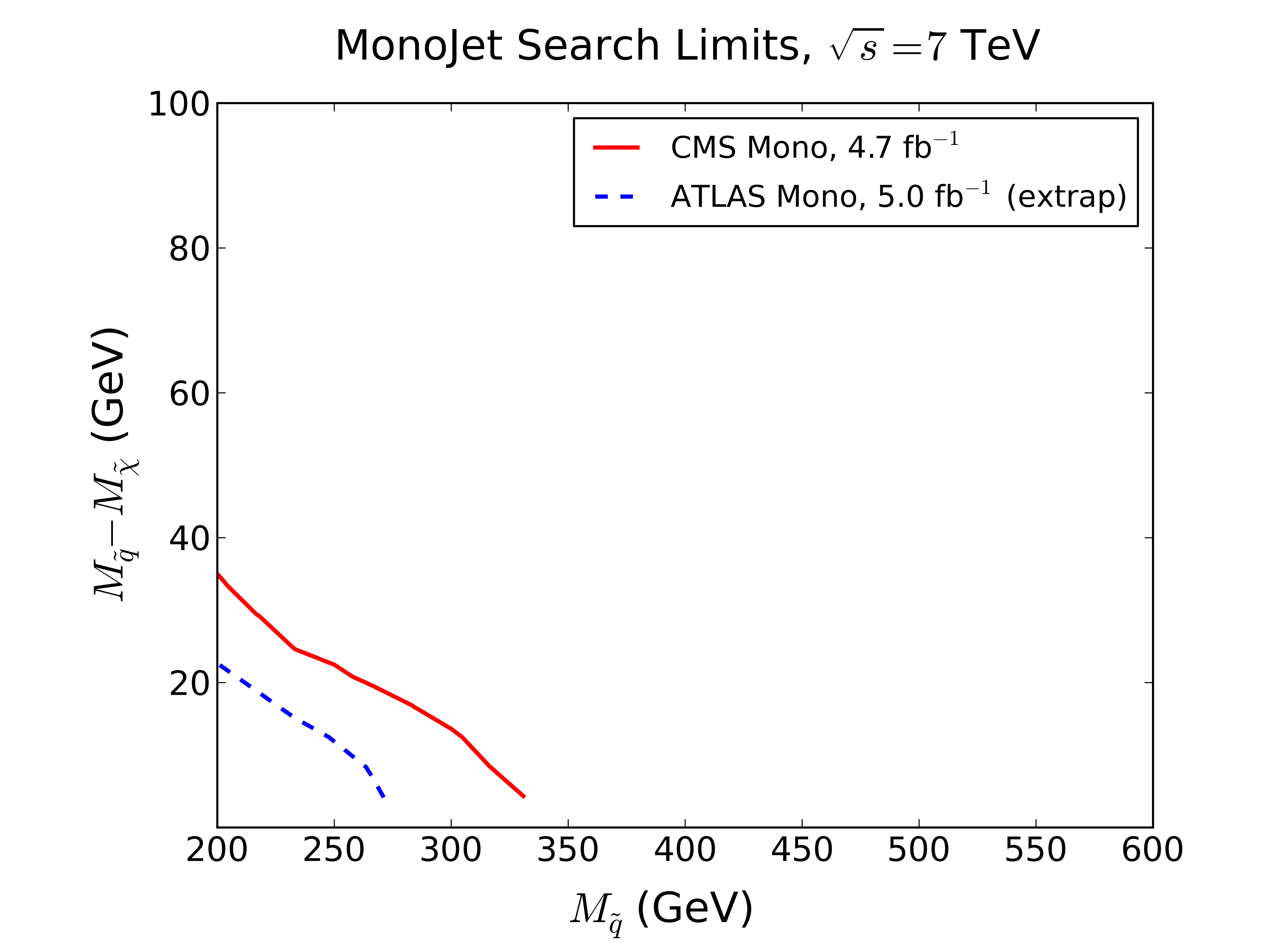}  
\includegraphics[scale=0.4]{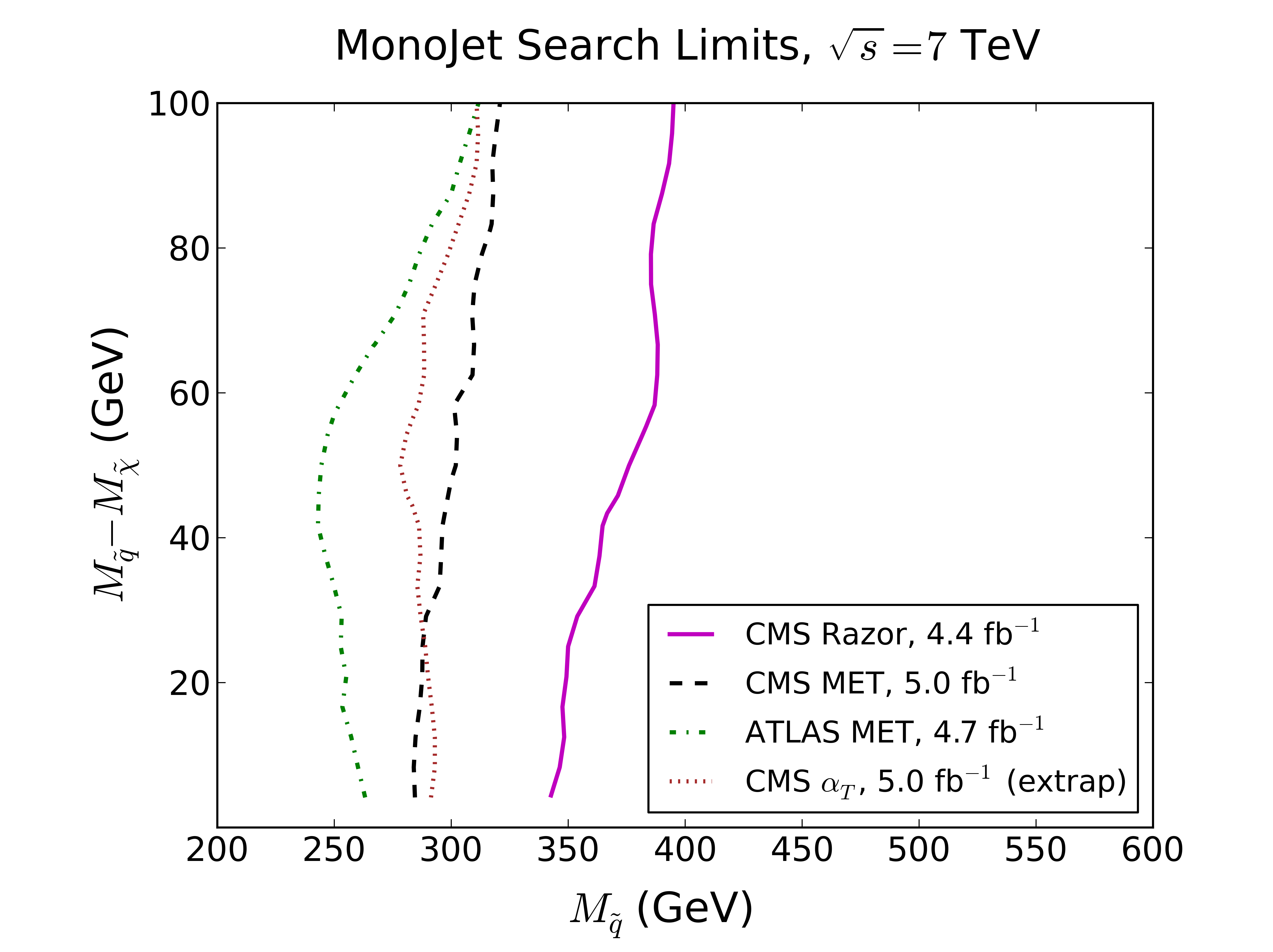}
\end{center} \vspace*{-0.4cm}
\caption{Limits at the 95\% confidence level for squark masses in the decoupled gluino scenario, Fig.\ref{fig:Spectrum}(a), for both monojet (left) and SUSY searches (right). The mass splitting between the squark and the LSP is varied between 1 and 100~GeV. \label{fig:SquarkLoopLimit}} 
\end{figure*}

In order to set the most stringent bound on the model parameter space we use all current SUSY hadronic searches from both ATLAS and CMS. In addition we apply the monojet searches from both experiments.

The motivation for using all searches is that there exists a range of strategies to look for SUSY at the LHC and it was 
not obvious which of these would be most productive for compressed spectra. The \textquoteleft vanilla' LHC searches are 
based around an effective mass and missing energy cut. Various signal regions 
are defined with different amounts and proportions of effective mass (or $H_T$) and missing energy. Both ATLAS \cite{ATLAS-
CONF-2012-033} and CMS \cite{CMS-PAS-SUS-11-004-B} have an analysis of this kind and they give similar results for mSugra. 
In addition, CMS has a number of \textquoteleft shape' based analyses using $\alpha_T$ \cite{Randall:
2008rw,Chatrchyan:2011zy}, $M_{T2}$ \cite{Lester:1999tx,CMS-PAS-SUS-12-002-B} and Razor \cite{Rogan:2010kb,CMS-PAS-
SUS-12-005-B}.

As ISR is vital for discovering SUSY in compressed spectra we would expect that the signal is dominated by the hard emission of a 
single parton from the initial state. Therefore, we also use the monojet analyses (ATLAS \cite{ATLAS-CONF-2011-096} and CMS \cite{CMS-Mono-B}) that are optimised for precisely this kind of signal.

All the searches were implemented within the RIVET \cite{Buckley:2010ar} analysis package. As we are only interested in jets, no experimental efficiencies apart from quality cuts were required. Momentum smearing of the jets, including adding mis-measured tails were included  \cite{Allanach:2011ej}\footnote{We thank the authors for providing the details of the jet mis-measurement constants in a private communication.}. However, these effects were found to have a negligible impact on all of the search efficiencies. All analyses were tested against any mSugra or simplified model bounds found in the original studies. In addition, where distributions and cut flows are available, these were also tested. Acceptances were reproduced to within 20\% for all analyses but were usually found to agree much better.

In order to set limits in a consistent way across all analyses, we use the 95\% confidence limit given by the Rolke Test \cite{Rolke:2000ij,Rolke:2004mj,Lundberg:2009iu} for the most constraining signal region in each analysis. To be able to fairly compare the analyses and gain a better understanding of the optimum search regions, if the limit is better than the limit expected to have been found, we take the expected limit. In this way we find more conservative limits than the official analyses.

Both CMS $\alpha_T$, and Razor searches use a combined likelihood over all signal regions to set a limit. We instead 
conservatively use only the best signal region as this allows a fairer comparison with the other searches. In addition, CMS Razor 
uses an unbinned likelihood that is impossible to reproduce as the data is not publicly available, however a 60 bin data 
set is available. We combined adjacent bins in this set to give a final search with 20 signal regions which is conservative compared 
to the official analysis.

At the time of release both the ATLAS monojet and CMS $\alpha_T$ only had analyses with 1~fb$^{-1}$ and 1.14~fb$^{-1}$. In order to give a fair comparison with the other analyses we extrapolated these searches to 5~fb$^{-1}$ by reducing statistical errors. This is a conservative approach as many systematic errors are also statistically limited and likely to be improved.

\begin{figure*}[ht!]
\begin{center}
\includegraphics[scale=0.4]{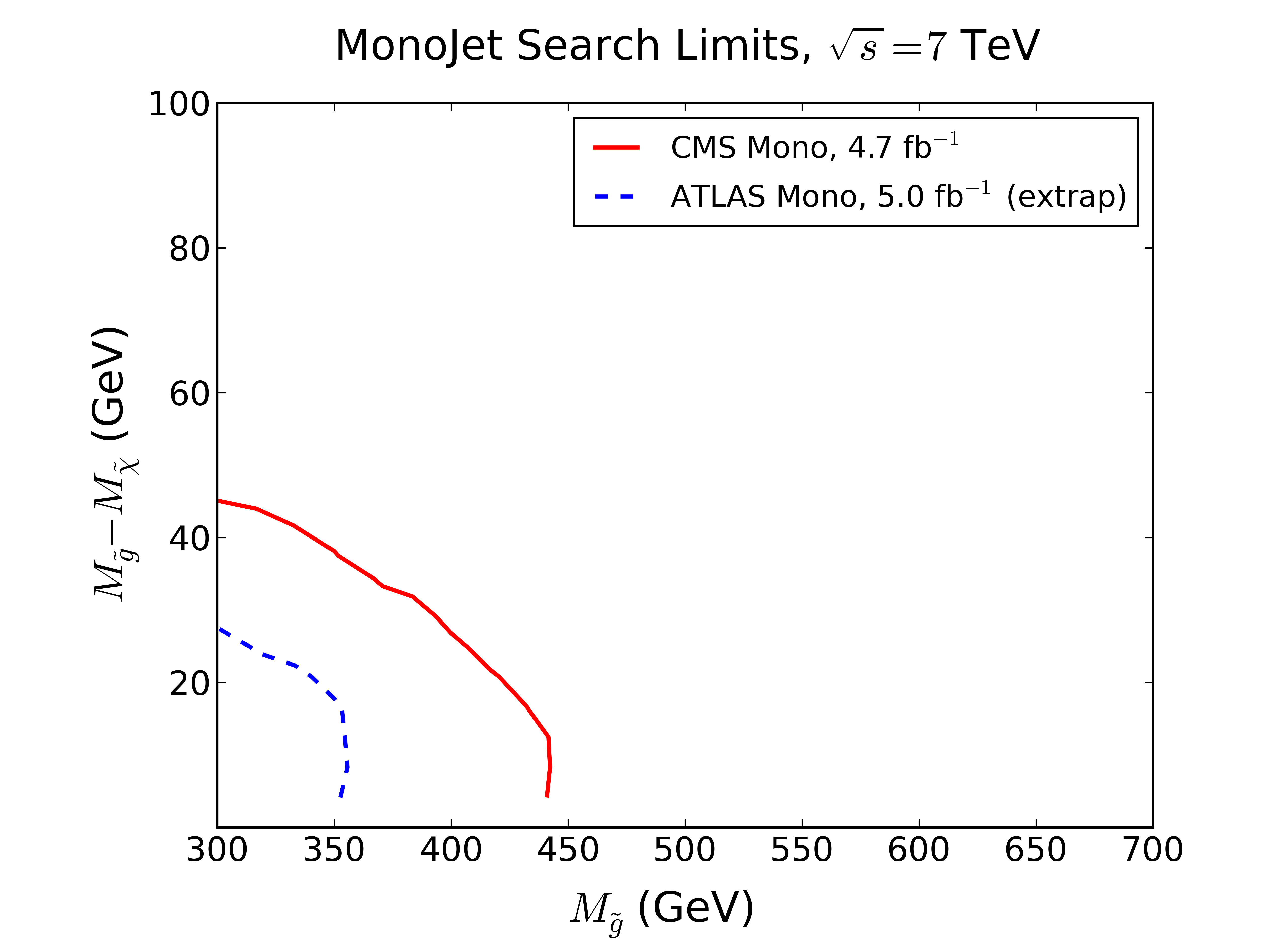}  
\includegraphics[scale=0.4]{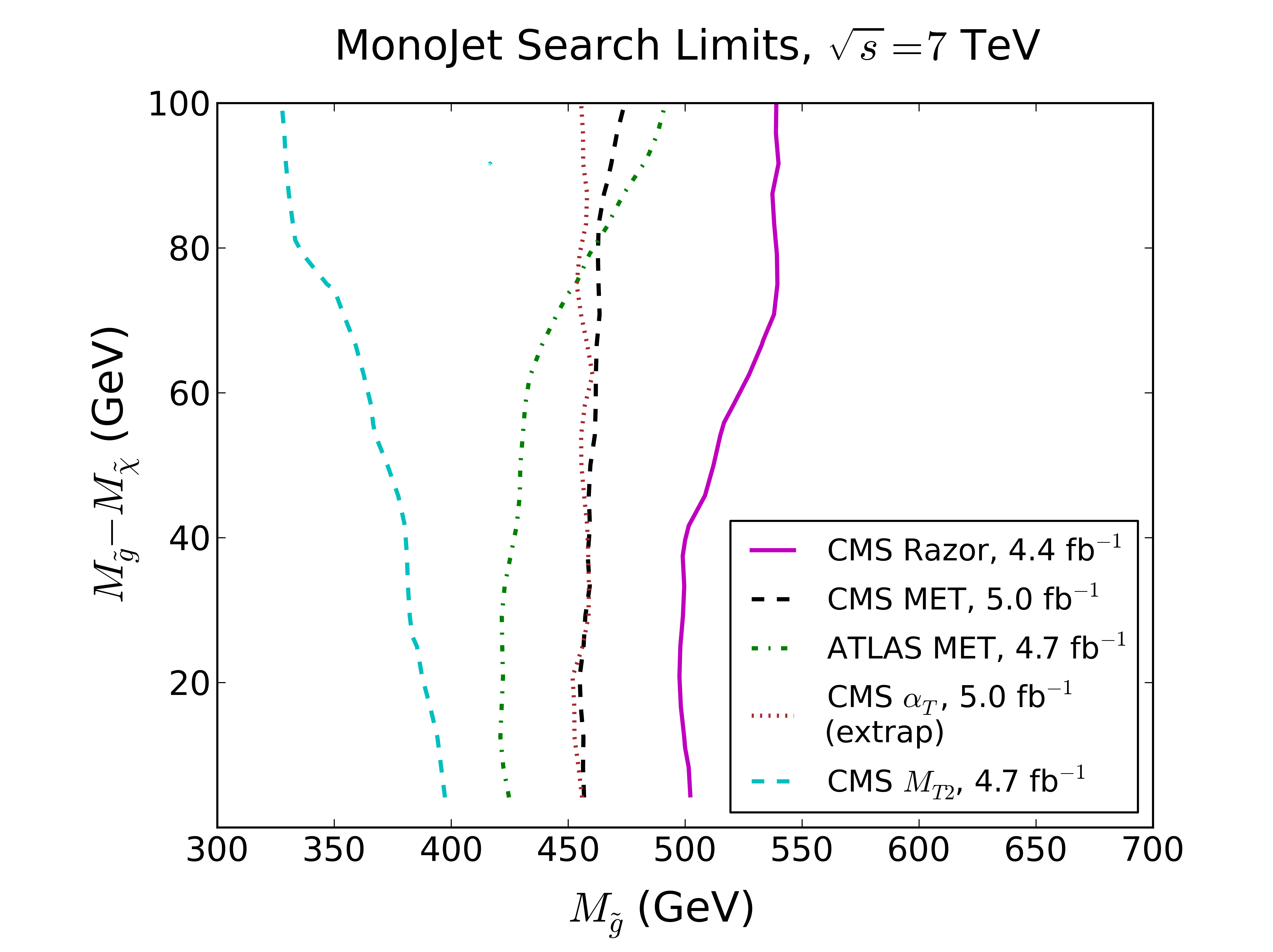}
\end{center} \vspace*{-0.4cm}
\caption{Limits at the 95\% confidence level for gluino mass in the decoupled squark scenario, Fig.\ref{fig:Spectrum}(b), for both monojet (left) and SUSY searches (right). The mass splitting between the gluino and the LSP is varied between 1 and 100~GeV.\label{fig:GluLoopLimit}} 
\end{figure*}

\begin{figure*}[ht!]
\begin{center}
\includegraphics[scale=0.4]{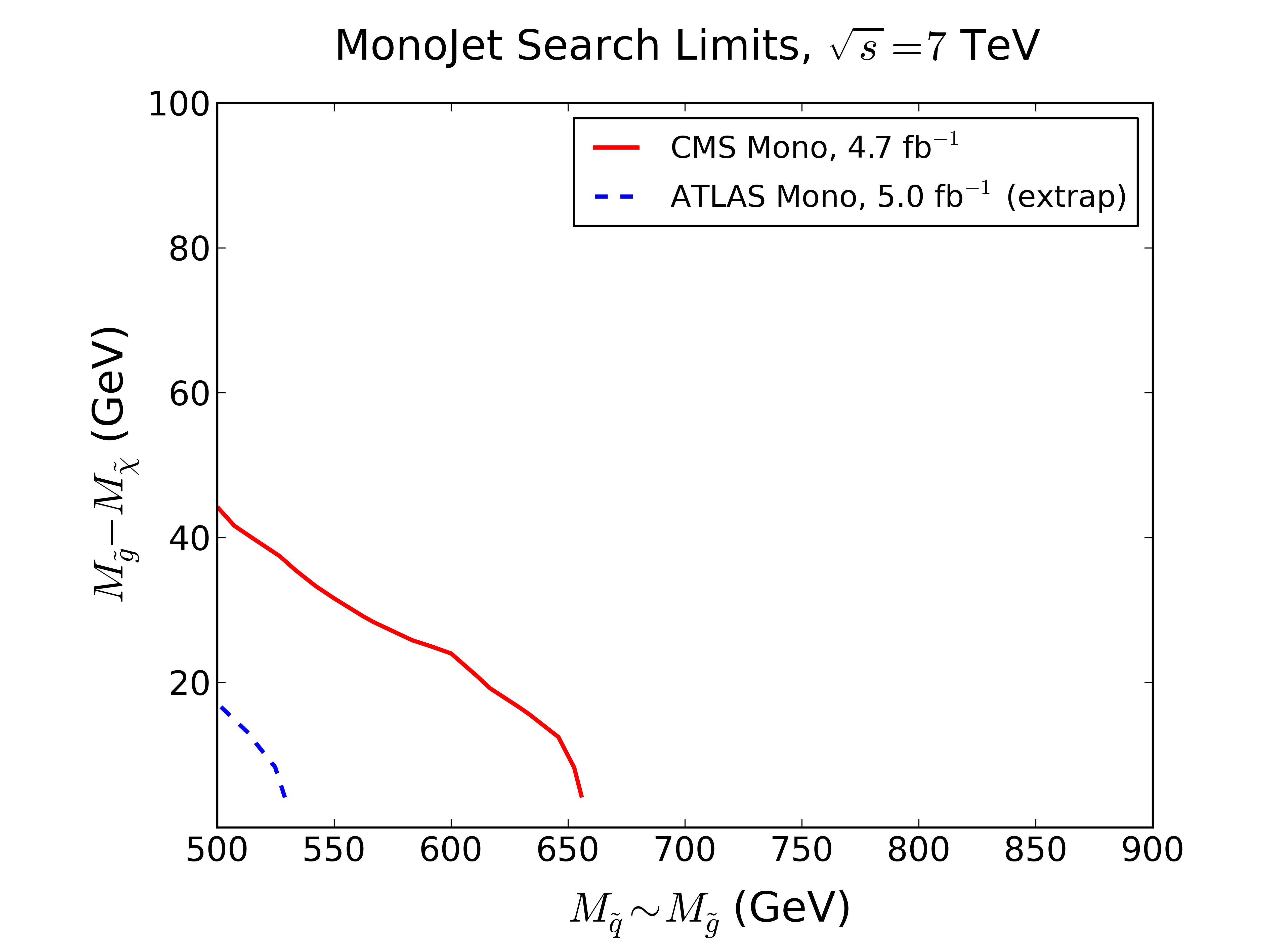}  
\includegraphics[scale=0.4]{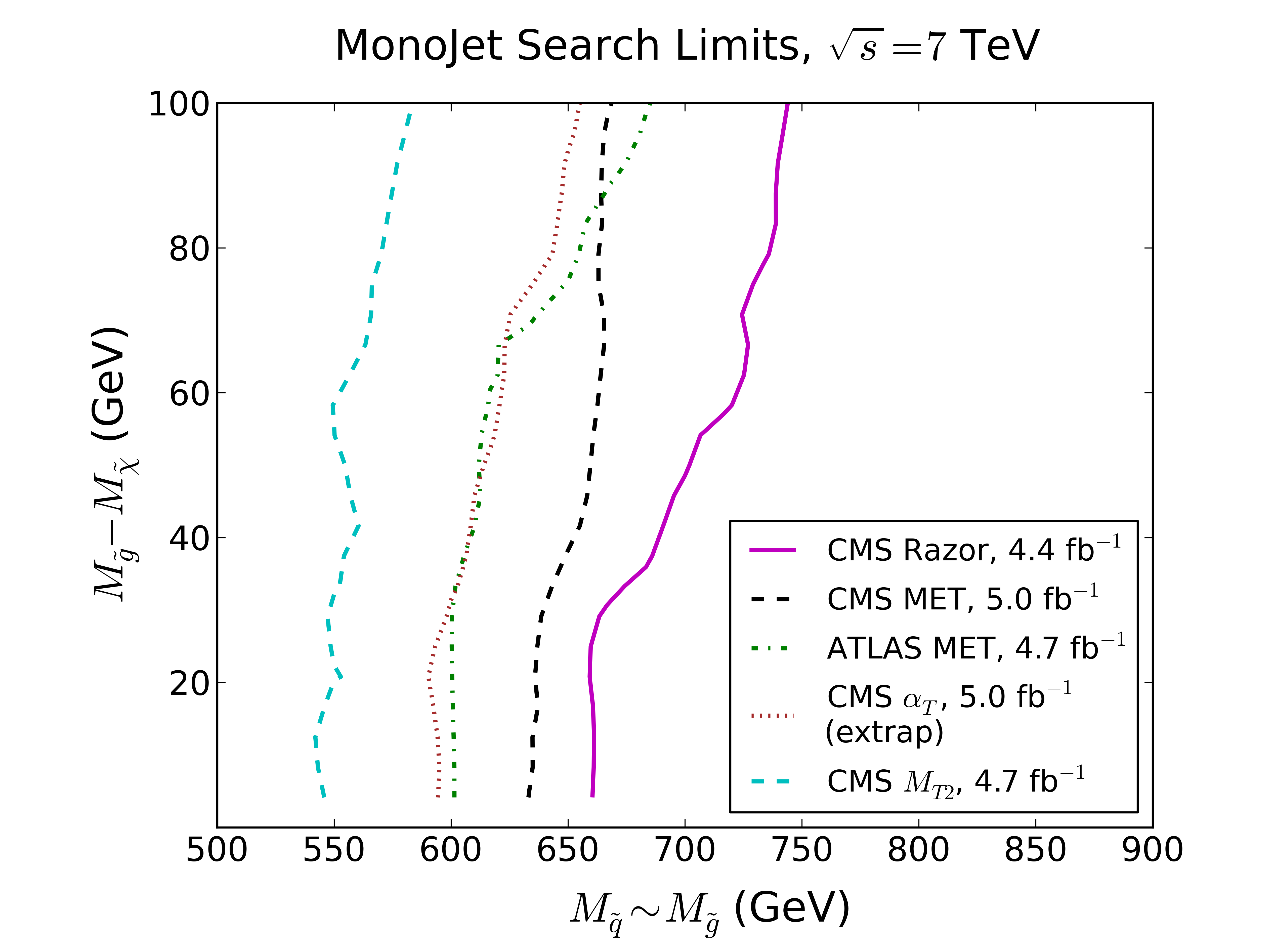}
\end{center} \vspace*{-0.4cm}
\caption{Limits at the 95\% confidence level for gluino masses in the equal mass squark-gluino scenario, Fig.\ref{fig:Spectrum}(c), for both monojet (left) and SUSY searches (right). The mass splitting between the gluino and the LSP is varied between 1 and 100~GeV whilst the squark mass is placed between the two.\label{fig:TotLoopLimit}} 
\end{figure*}

\section{Limits}
\label{sec:limits}

\begin{table*} \renewcommand{\arraystretch}{1.2}
\begin{center}
\begin{tabular}{|c||c|c|ccc|} \hline 
					& 				& Search Region			& 	 \multicolumn{3}{c|}{Lower Mass Bound (GeV)}  	 \\
 Search 				& $\mathcal{L}$ (fb$^{-1}$) 	& (given in source)			&Squark  & Gluino  	       & Squark $\sim$ Gluino \\ \hline\hline
 \underline{Monojet}				&				&				&	&		&\\
 ATLAS* \cite{ATLAS-CONF-2011-096}	& 	$5.0^{\dagger}$		&  High/veryHigh $p_T$		&$270$ & $350$ & $530$ \\ 
 \bf{CMS*}  \cite{CMS-Mono-B}		& 	4.7			&  $E_T^{\mathrm{miss}} > 400$	&\bf{340} & \bf{440} & \bf{650} \\ \hline\hline
 \underline{SUSY}					&				&  				&	&	&	\\
 ATLAS MET \cite{ATLAS-CONF-2012-033} 	& 	4.7			&	A' med/C med		&$260$ & $440$ & $600$  \\ 
  CMS $\alpha_T$ \cite{Chatrchyan:2011zy} & 	$5.0^{\dagger}$		&  Optimised $H_T$ bin 		&$290$ & $450$ & $600$  \\ 
  CMS MET \cite{CMS-PAS-SUS-11-004-B}	& 	5.0			&	A2			&$290$ & $450$ & $620$  \\ 
  CMS $M_{T2}$ \cite{CMS-PAS-SUS-12-002-B} & 	4.7			&	A/B			&- & - & $550$  \\ 
  \bf{CMS Razor} \cite{CMS-PAS-SUS-12-005-B} &	4.4			&	bHad($6_4+7_4+8_4+9_4$)	&  \bf{340} & \bf{500} & \bf{650}  \\  \hline
\end{tabular}
\caption{Comparison of the bounds on the mass of SUSY particles for the different searches employed at the LHC. The luminosity of the searches and the most constraining search region are also given (the search region names refer to those given in the original experimental papers). *\textit{The ATLAS and CMS monojet searches only give these bounds for mass differences $<5$~GeV. For larger mass splittings, the bounds become much weaker.} $^{\dagger}$\textit{The ATLAS monojet and CMS $\alpha_T$ searches are both extrapolated from $\sim$1~fb$^{-1}$ to give a more direct comparison.} \label{tab:Limits} }
\end{center}
\end{table*}

To calculate limits in our simplified models, cross sections are calculated to next-to-leading order including known next-to-leading-logarithmic corrections using NLL-Fast \cite{Beenakker:1996ch,Beenakker:2009ha,Beenakker:2011fu}. The theoretical uncertainty is calculated including the factorisation and renormalisation scale and PDF \cite{Nadolsky:2008zw} errors. In addition, we vary the matching scale and parton showers, Fig.\ref{fig:MGvsPyth}, and take the result with the least constraining bound. 

We find that in the decoupled gluino scenario with a quasi-degenerate LSP, Fig.\ref{fig:Spectrum}(a), we can set a bound on the first two generations of squarks of $M_{\tilde{q}}>340$~GeV. In this scenario we find that the best limit is set by both the CMS monojet and CMS Razor searches, Fig.\ref{fig:SquarkLineLimit}, whilst the other SUSY searches give slightly weaker limits of $M_{\tilde{q}}\gtrsim300$~GeV. All the limits along with the relevant search regions are given in Tab.\ref{tab:Limits}.

Interestingly, we also see that the extrapolated ATLAS monojet search provides a noticeably worse limit of $M_{\tilde{q}}>270$~GeV. Our analysis showed that the difference between the two monojet searches is primarily down to the second jet veto used in the ATLAS search which reduces the signal acceptance.

For all the SUSY analyses, we find the best search regions are where 
the proportion of missing energy compared to the total energy in the event is the highest. 
This is to be expected as the relevant events in our models that are expected to have a topology with a dominant jet that will recoil against the LSPs.

As we increase the mass splitting, $M_{\tilde{q}}-M_{LSP}$, we see that the monojet searches rapidly lose their effectiveness, Fig.
\ref{fig:SquarkLoopLimit}. This is due to the fact that both the ATLAS and CMS searches have a third jet veto. As the mass splitting 
is increased, the SUSY decays produce additional jets and these events are vetoed.

In contrast, the SUSY searches become more effective as the mass splitting is increased. For example, the CMS Razor limit 
improves from $M_{\tilde{q}}>340$~GeV when the mass splitting is 1~GeV to $M_{\tilde{q}}>400$~GeV when the mass splitting is 
100~GeV. This is because the SUSY searches often allow any number of additional jets and include the extra radiation in the 
search variable. 

In the decoupled squark scenario, Fig.\ref{fig:Spectrum}(b), we find that we can set a bound on the gluino mass, $M_{\tilde{g}}
>500$~GeV, Fig.\ref{fig:GluLoopLimit}, and CMS Razor sets the best limit.
Finally, in the equal mass squark gluino model, Fig.\ref{fig:Spectrum}(b), we find a limit of $M_{\tilde{q}} \sim M_{\tilde{g}}
>650$~GeV which is set both by CMS Razor and the CMS monojet search, Fig.\ref{fig:TotLoopLimit}. As before, the limit in the 
monojet search is reduced as we increase the mass splitting but the majority of the SUSY limits actually improve and for 100~GeV 
mass splitting, CMS Razor gives a limit, $M_{\tilde{g}}>700$~GeV.

\section{Conclusions}
\label{sec:conclusions}

We have set limits in simplified SUSY models with compressed spectra. We find a limit on the first two generations of squarks, $M_{\tilde{q}}>340$~GeV, in a model with a decoupled gluino. With decoupled squarks we find a limit on the gluino mass of, $M_{\tilde{g}}>500$~GeV. With equal mass gluinos and squarks we find a limit of $M_{\tilde{q}} \sim M_{\tilde{g}}>650$~GeV.

We would like to comment that the results for the decoupled squark and equal mass squark-gluino models are in good agreement for the ATLAS SUSY search with those previously presented in \cite{LeCompte:2011cn,LeCompte:2011fh}.

%

\section{Acknowledgements}
\noindent We would especially like to thank Stefan Prestel for his help adapting the Pythia 8 matching algorithm. In addition we 
would like to acknowledge useful discussions with John Conley, Krysztof Rolbiecki, Daniel Wiesler, Johan Alwall and Maurizio 
Pierini. This work has been supported in part by the Helmholtz Alliance `Physics at the Terascale' and the DFG SFB/TR9 ``Computational Particle
Physics''. HD was supported by BMBF Verbundprojekt HEP-Theorie� under the contract 0509PDE.

\bibliographystyle{eplbib}
\bibliography{ISR_SUSY}

\end{document}